# Dynamic functional connectivity: why the controversy?

Diego Vidaurre[a, b]

a - Center of Functionally Integrative Neuroscience, Department of Clinical Medicine, Aarhus University, Aarhus, Denmark
b - Oxford Centre for Human Brain Activity, Psychiatry Department, Oxford University, Oxford, UK

**Abstract**: In principle, dynamic functional connectivity in fMRI is just a statistical measure. A passer-by might think it to be a specialist topic, but it continues to attract widespread attention and spark controversy. Why?

Functional connectivity (FC) is defined as some statistical relationship between the activity of two or more brain areas (Smith, et al., 2013; Friston, 2011). This relationship is typically quantified applying linear measures between each pair of areas such as covariance or Pearson's correlation, but nonlinear measures are also possible (Tedeschi, et al., 2005). The most straightforward measure of FC is the so-called static FC, which considers purely instantaneous relations between the time series; for instance, if we have two time series *x* and *y* with *T* time points each, the covariance $x' y / (T-1)$ would be a measure of static FC. Dynamic FC refers to the temporal aspects of FC (Lurie, et al., 2020), but a precise definition is often lacking. Here, I use a broad definition: dynamic FC is any cross-area information that is not captured by static FC. Considering the BOLD signal's temporal information, this general definition of dynamic FC spans, as depicted in **Figure 1**, two different aspects. The first is time-varying, instantaneous FC, in the sense of within-session modulations in e.g. covariance or correlation (Hutchison, et al., 2013). The second is non-instantaneous FC dynamics in the sense of a parametric relation between the *now* of one time series and the *later* of the other, as captured for example by an autocorrelation, and autocovariance, or an autoregressive model (Harrison, et al., 2003). So, to recapitulate, I consider (linear or nonlinear) dynamic FC as a broad term encompassing both time-varying FC and non-instantaneous FC dynamics.

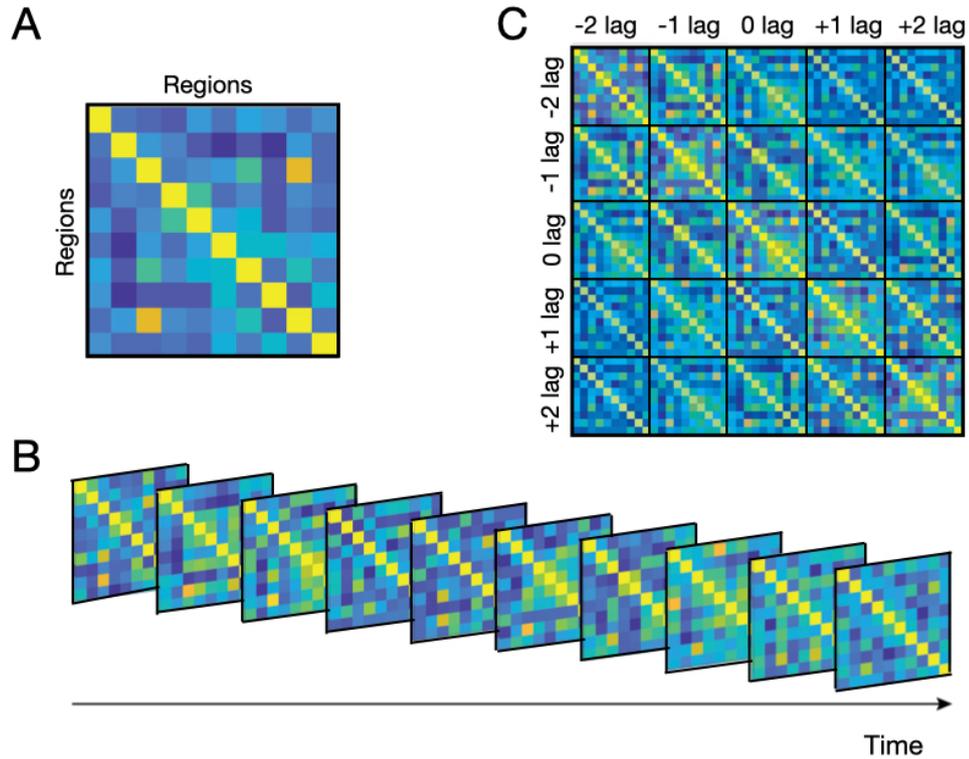

**Figure 1**. Examples of static (**A**) and dynamic (**B**,**C**) estimates of functional connectivity (FC); panel **B** reflects the time-varying aspect of dynamic FC, while panel **C** is an autocovariance matrix reflecting the non-instantaneous FC aspect.

The controversy I refer to concerns the time-varying aspect, and much of it revolves around two arguments: that other models that are not time-varying fit the data better (Liegeois, et al., 2017), and that FC is stable and varies little in comparison to other axes of variability (Gratton, et al., 2018). There are also practical concerns about statistical variability and spurious results (Laumann, et al., 2016), but these relate more to specific methods than to time-varying FC as a whole; therefore, they can be addressed technically, as discussed elsewhere (Vidaurre, et al., 2017). Finally, there is a broader debate about the aetiology of time-varying FC in particular and spontaneous fluctuations of the BOLD signal in general, and the extent to which these reflects higher-order cognition (Laumann & Snyder, 2021), but this will not be discussed either given the focus of this letter on the modelling facet.

I feel part of the debate stems from a belief that time-varying and non-instantaneous dynamics are *generally* distinct and separable components. They can be, but only if we assume that the generative model of the signal is of a particular form. The problem is that we are very far from having a realistic generative model of real fMRI data, where realistic means a model that reasonably captures the biological processes under the hood. In the absence of such model, estimates of either the first or the second type of information are just complementary statistics with overlapping information. That is, there is no universal statistical model for such a complex object as the brain or many other aspects of nature (Box, 1976).

To better understand this point, let us assume we have pseudo-fMRI signals $X_t$ (where t indexes time points) sampled from a multivariate Gaussian distribution with mean $X_{t-1} A$ and covariance $\Sigma_k$ defined over a number of areas or voxels. Here, $A$ is a (number of areas by number of areas) matrix of autoregressive coefficients that endows the signal with cross-area temporal dynamics, whereas the covariance $\Sigma_k$ models instantaneous, structured noise. This model assumes that there are different covariance matrices available, indexed by $k$: $\Sigma = \{ \Sigma_k,$

$k=1,…,K$ } —so that, within a session, the signal can change between different patterns of structured noise. Together, this constitutes a latent-state model that can be used to sample signals with chaotic behaviour that mimic some of the characteristics of real fMRI data. In this model, $Σ$ captures time-varying, instantaneous FC, whereas $A$ encodes non-instantaneous FC dynamics —so both contain dynamics as per the definitions above. Given data generated from this model, these two types of information can be estimated and disentangled by fitting the model parameters.

Now, let us consider that, instead of jointly estimating the parameters $A$ and $Σ$, we fit two simpler models to the synthetic data. One is a standard Hidden Markov model with estimated covariance matrices $Σ'$ (Vidaurre, et al., 2017; Vidaurre, et al., 2018), and the other is an autoregressive model $A'$ (Harrison, et al., 2003); the former captures time-varying instantaneous FC, and the latter models non-instantaneous FC dynamics. Because these models are not complete with respect to the generative model, their respective estimates $Σ'$ and $A'$ will differ from the ground truth $Σ$ and $A$. But this does not mean that these estimates are non-informative. In fact, they represent features of the data at a certain level of abstraction, and contain information that can be leveraged to address different research questions. I stress that I focus on specific models (the autoregressive and the hidden Markov model) just to make the point more concrete, but similar arguments could be made with alternative methods.

Back to actual fMRI, it is possible to fit the above model to real data, which, unlike the standard Hidden Markov and the autoregressive model, captures both types of information. We can then apply some statistical test to assess whether the noise covariance changes within a session. If this test is significant, we can take this as safe evidence of the existence of time-varying FC *given this model*. This diagnostic approach is appropriate to decide whether this is a reasonable candidate model for further analyses or we can simplify it. But can we use this test to make a *general* claim about the existence of time-varying FC? Unfortunately not, because the brain is nothing like this model and time-varying FC is not a brain mechanism, but a statistical measure we —as researchers— invented as a convenient proxy of neural dynamics. On these grounds, we are free to choose if we represent the data using a time-varying FC model (e.g. based on $Σ$) or a linear dynamical system (e.g. based on a single set of parameters $A$), because neither of them is more correct than the other. Often, this also applies to questions about the "right" choice of hyperparameters in these models; in the absence of a compelling hypothesis of the underlying generative model, there is no right or wrong in absolute terms.

Still, could we ask which of the two types of information is a better description of the data, in objective terms, such as explained variance or number of parameters? Not with absolute objectivity, because there are numerous models that may capture these types of information, and numerous metrics that could be used to compare these models. Although quantitative model evaluation is considered a scientific gold standard, and there are of course reasonable metrics available, the choice of model and metric is ultimately subjective.

A more modest, but arguably more reasonable, question we can ask is: which model is more useful for the specific question and objective of our analysis? Whereas a model of non-instantaneous FC dynamics might reflect better some aspects of the data, such as its spectral properties, a time-varying FC method might be more useful when the focus is on temporal changes. For instance, Baldassano and colleagues trained Hidden Markov models on data acquired while subjects watched a film, and, fixing the state parameters, computed the most plausible sequence of state activations while the subjects internally recollected the story (Baldassano, et al., 2017). This way, they were able to uncover differences in the time scale of the neural dynamics between brain areas and also between watching versus recollection. In another example, Stevner and colleagues also used Hidden Markov models to offer, in a data-driven way, a refined perspective of the sleep cycle (Stevner, et al., 2019), which is known to go through a set of discrete stages. In these cases, a model that captures non-instantaneous cross-region statistics using a single set of parameters, instead of a time-varying model with identifiable temporal fluctuations, would have been less straightforward to interrogate in relation to these two scientific questions.

Another aspect of the controversy is about explained variance. If we apply principal component analysis to our data and sort the components sorted by their explained variance, then the chance that we find an interesting effect is —in the absence of any other information— statistically greater in the higher-order than in the lower-order components, simply because there is more information in the former. But, critically, this is not guaranteed. We have plenty of examples across disciplines where the interesting effect occurs in low-order components. For instance, in genetic epidemiology the first principal components of variation in a population do not normally relate to a disease of interest, but to ancestry and other aspects of population structure. For this reason, genome-wide association studies often include the first principal components as confounding covariates (Price, et al., 2006). For analogous reasons, the fact that between-subject variability in static FC is considerably larger than time-varying, within-session variability (Gratton, et al., 2018) does not mean that the latter is irrelevant; and it is possible, in certain cases, that the smaller within-session variability is more predictive of behavioural effects than the larger static FC variability (Vidaurre, et al., 2021; Liégeois, et al., 2019). Sometimes, indeed, the devil is in the detail.

In summary, time-varying FC is, at the end of the day, just a statistical measure. Whether the data really affords that level of description and how well we can estimate it depends on both the data and our methods of analysis (Ahrends, et al., 2022); and whether it makes sense or not to use depends on the scientific question at hand. Arguments about whether time-varying FC is a truthful representation of brain activity and neural communication, as opposed to a stationary model or any other metric, are, in the author's opinion, a mere strawman.

**Acknowledgements**: DV is supported by a Novo Nordisk Foundation Emerging Investigator Fellowship (NNF19OC-0054895) and an ERC Starting Grant (ERC-StG-2019-850404). I also thank the Wellcome Trust for support (106183/Z/14/Z, 215573/Z/19/Z). This research was funded in part by the Wellcome Trust (215573/Z/19/Z). For the purpose of Open Access, the author has applied a CC BY public copyright licence to any Author Accepted Manuscript version arising from this submission. Finally, I would like to thank Dan Bang and Christine Ahrends for their useful comments.